# Topological Invariants in Point Group Symmetric Photonic Topological Insulators

Revised 7/21/2014  06:11:00

Xiao-Dong Chen, Zi-Lan Deng, Wen-Jie Chen, Jia-Rong Wang, and Jian-Wen Dong*
*State Key Laboratory of Optoelectronic Materials and Technologies & School of Physics and Engineering*
*Sun Yat-Sen University, Guangzhou, China*
e-mail address: dongjwen@mail.sysu.edu.cn

We proposed a group-theory method to calculate topological invariant in bi-isotropic photonic crystals invariant under crystallographic point group symmetries. Spin Chern number has been evaluated by the eigenvalues of rotation operators at high symmetry k-points after the pseudo-spin polarized fields are retrieved. Topological characters of photonic edge states and photonic band gaps can be well predicted by total spin Chern number. Nontrivial phase transition is found in large magnetoelectric coupling due to the jump of total spin Chern number. Light transport is also issued at the ε/μ mismatching boundary between air and the bi-isotropic photonic crystal. This finding presents the relationship between group symmetry and photonic topological systems, which enables the design of photonic nontrivial states in a rational manner.

Group theory has been widely used in solid state physics, linear optics and quantum physics. Identifying symmetries in condense matter systems can simplify its complexity and subsequently predigest physics solutions. Due to the mathematical analogy between photonic and electronic systems, symmetry in photonic crystals also plays an important role in optical responses. For example, Fano resonances were found at k≈0 in macroscopic photonic crystal slab due to symmetry mismatching [1-3]. Zero-refractive-index photonic crystals were achieved by a triple accidental degeneracy state [4-6]. Bidirectionality and nonreciprocity in magnetic photonic crystals have been studied through magnetic group theory [7].

On the other hand, topological order has been actively explored in many photonic systems [8-25]. Topologically protected edge states have been realized in magnetic photonic crystals [8-11], resonating lattices [12], and evanescently coupled helical waveguides [13]. Photonic analogue of topological insulators [26, 27], i.e., photonic topological insulators (PTI), have also been accomplished [14-18]. One is coupled resonator optical waveguides in which clockwise (anticlockwise) waveguide mode is regarded as pseudo spin-up (spin-down) state [14-16]. Another is ε/μ-matching photonic crystals with bianisotropic coupling in which photonic spin pair is formed by linear combinations of transverse magnetic and transverse electric modes [17, 18]. One of intriguing questions is to address the connection between the topology of nontrivial photonic systems and the group theory, so as to make it accessible in a simple way.

In this paper, we predigest the determination of topological invariants by group theory when bi-isotropic photonic crystals possess point group symmetries. Spin Chern number of $C_n$-invariant bi-isotropic photonic crystal can be accurately calculated, up to a multiple of n, by evaluating the eigenvalues of symmetry operators at high symmetry k-points in the Brillouin zone (BZ). Total spin Chern number is then used to point out whether the photonic band gap is nontrivial or not and to characterize the spin-polarized edge states. Topological phase transition is realized when the magnetoelectric coupling is large enough to induce the jump of total spin Chern number and mode exchange. Light transport at the edge supporting pseudo-spin polarized state will be also discussed in a ε/μ matching or mismatching systems.

Consider a bi-isotropic medium with the reciprocal constitutive relations of $\mathbf{D} = \varepsilon_0 \varepsilon_r \mathbf{E} + \boldsymbol{\zeta} \mathbf{H}$ and $\mathbf{B} = \mu_0 \mu_r \mathbf{H} + \boldsymbol{\zeta} \mathbf{E}$. Here, $\boldsymbol{\zeta}$ is the magnetoelectric coefficient tensor with nonzero elements of $\zeta_{12} = \zeta_{21}^* = -i\zeta_0/c$. $\varepsilon_r$ and $\mu_r = \varepsilon_r/\alpha$ are the relative permittivity and permeability, where $\alpha$ is a constant in the whole space (so-called ε/μ matching condition). Introducing the pseudo-fields $\mathbf{P} \equiv (\mathbf{P}^+, \mathbf{P}^-) \equiv (\sqrt{\alpha\varepsilon_0}\mathbf{E} + \sqrt{\mu_0}\mathbf{H}, \sqrt{\alpha\varepsilon_0}\mathbf{E} - \sqrt{\mu_0}\mathbf{H})$, the full-vector Maxwell equations can be reformulated:

$$\nabla \times \begin{pmatrix} P_x^- \\ P_y^- \\ P_z^+ \end{pmatrix} = -i\frac{\omega}{c} \begin{pmatrix} \mu_r\sqrt{\alpha} & i\zeta_0 & \\ -i\zeta_0 & \mu_r\sqrt{\alpha} & \\ & & -\mu_r\sqrt{\alpha} \end{pmatrix} \begin{pmatrix} P_x^- \\ P_y^- \\ P_z^+ \end{pmatrix} \quad (1)$$

$$\nabla \times \begin{pmatrix} P_x^+ \\ P_y^+ \\ P_z^- \end{pmatrix} = i\frac{\omega}{c} \begin{pmatrix} \mu_r\sqrt{\alpha} & -i\zeta_0 & \\ i\zeta_0 & \mu_r\sqrt{\alpha} & \\ & & -\mu_r\sqrt{\alpha} \end{pmatrix} \begin{pmatrix} P_x^+ \\ P_y^+ \\ P_z^- \end{pmatrix} \quad (2)$$

As Eq. (1) is decoupled from Eq. (2), we can define spin-up polarization of $(0,0,P_z^+, P_x^-, P_y^-, 0)^T$ and spin-down polarization of $(P_x^+, P_y^+, 0, 0, 0, P_z^-)^T$. When the magnetoelectric coupling in bi-isotropic photonic crystals is large enough, photonic spin-degenerated band gap will open. The topology of the band gap is characterized with the topological invariant of bulk bands.

Chern number is the representative topological invariant of pseudo-spin polarization in bi-isotropic photonic crystals. In general, the Chern number of the $m$th spin-up polarized bulk band can be are calculated with the determinant of closed loop integral of Berry connection along the boundary of BZ,

$$\exp\left(i 2\pi C_m^+\right) = \det\left\{Path \exp\left[\oint_{BZ} \left\langle P_{z;m}^+(\mathbf{k})\middle|\nabla_k\middle|P_{z;m'}^+(\mathbf{k})\right\rangle d\mathbf{k}\right]\right\} \quad (3)$$

where "det" means determinant, "$Path$" means path ordered and $\left\langle P_{z;m}^+(\mathbf{k})\middle|\nabla_k\middle|P_{z;m'}^+(\mathbf{k})\right\rangle$ is Berry connection. The subscript "m" stands for the $m$th bulk band, the subscript "z" stands for the z-component of $\sqrt{\alpha\varepsilon_0}\mathbf{E} + \sqrt{\mu_0}\mathbf{H}$, and the superscript "+" stands for the spin-up polarization. According to time-reversal symmetry, the Chern number of the $m$th spin-down polarized bulk band is $C_m^- = -C_m^+$, and then the spin Chern number is simplified to $C_s^m = (C_m^+ - C_m^-)/2 = C_m^+$. However, the photonic Bloch functions have no analytical solutions and the numerical simulations are always phase discontinued. In such case, the calculation of the spin Chern number by employing Eq. (3) is extremely difficult. Alternatively, if the photonic system is $C_n$-invariant, Berry connections at the $C_n$-rotation connected k-points are equal. So one just need to integrate the Berry connection along the boundary of the irreducible BZ and get one-$n$th of the Berry phase (i.e., $2\pi C_s/n$). Due to the periodicity of irreducible BZ and the $C_n$-invariance, the determinant of the closed loop integral can be evaluated with the determinant of sewing matrix at high symmetric reciprocal k-points [28]. As a result, the spin Chern number can be further obtained, up to a multiple of n, by the eigenvalues of rotation operators at high symmetry k-points, regardless of the photonic Bloch wave function at all other k-points. For example, one can determine the spin Chern number of a hexagonal bi-isotropic photonic crystal as follow [28],

$$e^{i\left(\frac{2\pi}{6}C_s^m\right)} = e^{i\left(\frac{2\pi}{6}C_m^+\right)} = \eta_m^+(\Gamma)\theta_m^+(K)\gamma_m^+(M) \quad (4)$$

where $\eta^+(\Gamma)$, $\theta^+(K)$ and $\gamma^+(M)$ represent the eigenvalues of $C_6$, $C_3$ and $C_2$ rotation operators on the spin-up modes at $\Gamma$, K and M points, respectively. Note that this method has obvious superiority to reduce the complexity of spin Chern number calculation, compared to the way of integrating Berry connection along the boundary of BZ.

In order to illustrate how to use the group-theory-based calculation method, we investigate the hexagonal bi-isotropic photonic crystal [inset of Fig. 1(a)]. The radius of the green rod is $0.3a$, where $a$ is the lattice constant. As the material configurations just change the width of photonic band gap, we will assume the "air" configuration: the unit cell consists of the bi-isotropic rod $(1,1,\zeta_{air})$ embedded in the anti-biisotropic background $(1,1,-\zeta_{air})$. In experimental realization, "anti" configuration may be fulfilled by flipping the positions of perfect electric and magnetic conductors in different domains in waveguides. Note that the band structure of other material configurations $(\varepsilon_r, \mu_r, \zeta)$ can be obtained by scaling down (or up) that of the "air" configuration with $\zeta_{air} = \zeta/\sqrt{\varepsilon_r\mu_r}$ as long as the system has same unit-cell configuration [29].

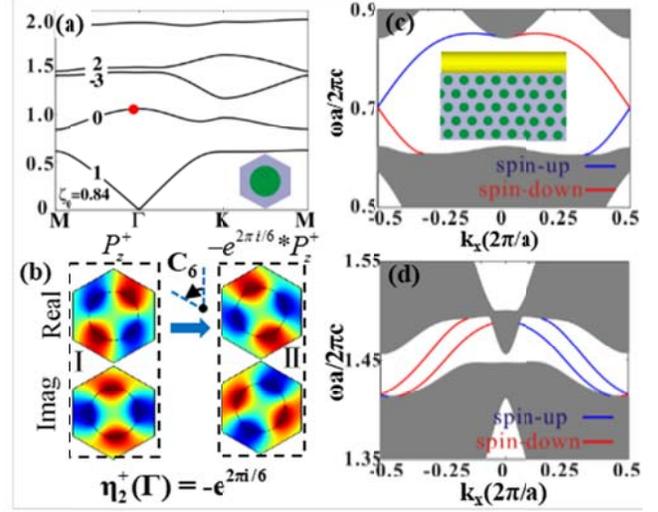

FIG. 1 (color online). Spin Chern numbers ($C_s$) and spin-polarized edge state in hexagonal bi-isotropic photonic crystals. (a) Bulk bands with $C_s$ when $\zeta_0 = 0.84$. The unit-cell consists of the green rod $(1,1,\zeta_{air})$ and the purple background $(1,1,-\zeta_{air})$. (b) Determination of $C_6$ eigenvalue for spin-up polarized state at $\Gamma$ point [red circle in (a)]. (c) & (d) Simulated spin-up (blue) and spin-down (red) edge states, of which the topological characters is accurately predicted from the total spin Chern number simply calculated by group theory.

Figure 1(a) shows the bulk band structure with the spin Chern number for $\zeta_0 = 0.84$, calculated by adding bi-isotropic constitutive equations into COMSOL Multiphysics. There are four band gaps below $\omega a/2\pi c = 2$. Figure 1(b) illustrates how to determine the $C_6$ eigenvalues for spin-up mode at $\Gamma$ point of the second band (m = 2), $\eta_2^+(\Gamma) = -e^{2\pi i/6}$. The complex $P_z^+$ fields II are equal to the fields I multiplying by the complex number $-e^{2\pi i/6}$, and it can also obtained by simply rotating the fields I anti-clockwise by 60°. By similar procedure, we have $\gamma_2^+(M) = 1$ and $\theta_2^+(K) = e^{2\pi i/3}$ [29]. Consequently, the spin Chern number of the second band is determined by Eq. (4) to $C_s^2 = 0$. Table 1 lists the $C_n$ eigenvalues of the spin-up modes at high symmetry k-points of the four lowest bands, and the spin Chern number of each band is 1, 0, -3, and 2, respectively. Note that the one should retrieve correct spin polarized states before calculating spin Chern number [29].

Topology classifications of the four lowest band gaps are characterized by the total spin Chern number $\sum C_s^m$, as listed in Table 1. The three lowest band gaps are topologically nontrivial due to non-zero total spin Chern number, while the fourth gap is trivial as $\sum_1^4 C_s^m = 0$. In particular, the first two gaps are topologically distinct from the third gap because of the different sign of the total spin Chern number. The topological distinction can be revealed from the spin polarized edge states. We constructed an edge along the ΓK direction between the photonic crystal and a trivial gap material with homogeneous anisotropic constitutive parameters of $\boldsymbol{\varepsilon}_r = \boldsymbol{\mu}_r = \text{diag}(1,1,-10000)$ for simplicity. One may fill with another bi-isotropic photonic crystal with trivial bandgap in experiment. For the first two gaps, only one gapless edge state spans the whole band gap. For the first gap, the group velocities of spin-up edge states point to the +x direction while those of spin-down edge states to the -x direction (Fig. 1(c)). This is similar to chiral edge states in quantum Hall systems with opposite external fields, and it is consistent with total spin Chern number of $\sum C_s^m = 1$. However, the total spin Chern number is -2 for the third gap. It means that there are two gapless edge states but with directions opposite to those in the first gap [30], which is verified in Fig. 1(d). These topological features should not change for different edge morphologies or edge directions as they are determined by the topological invariant [29]. The zero total spin Chern number for the fourth gap indicates its trivial properties, and thus the spin edge states are gapped and form a close loop instead of gapless connection.

Table 1 $C_n$ eigenvalues and spin Chern number of the four lowest bands

| Index | $\eta_m^+(\Gamma)$ | $\theta_m^+(K)$ | $\gamma_m^+(M)$ | $C_s^m$ | $\sum C_s^m$ |
|---|---|---|---|---|---|
| m=4 | $e^{-2\pi i/6}$ | 1 | -1 | 2 | 0 |
| m=3 | -1 | 1 | 1 | -3 | -2 |
| m=2 | $-e^{2\pi i/6}$ | $e^{2\pi i/3}$ | 1 | 0 | 1 |
| m=1 | 1 | $e^{-2\pi i/3}$ | -1 | 1 | 1 |

Figure 2 plots the gap evolution as a function of $\zeta_0$ to investigate the phase transition. Trivial gaps with $\sum C_s^m = 0$ are marked by red color while nontrivial gaps are marked by blue or cyan colors. One can see that the first two gaps will open at a certain value of $\zeta_0$ and keep a fixed total spin Chern number, showing its behaviors as photonic topological insulator and the support of robust helical edge states. However, the case is more complicated for the other two gaps. For the third gap, it opens at $\zeta_0 = 0.5$ and closes at $\zeta_0 = 0.816$. Between this range (red), the total spin Chern number is zero, so that the crystal is photonic ordinary insulator in which the edge states will suffer from backscattering. The crystal will turn into photonic topological insulator after it experiences phase transition at $\zeta_0 = 0.816$ or $\zeta_0 = 0.858$. Note that each phase in the gap map is an isolated "island", and a phase cannot be adiabatically connected to its neighboring phase without closing and reopening the gap.

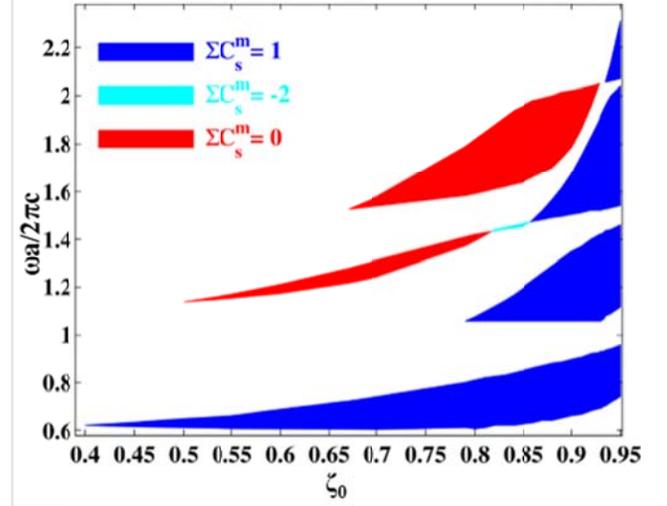

FIG. 2 (color online). Topological phase transitions of four lowest gaps. Blue and cyan for nontrivial gaps with $\sum C_s^m = 1$ and -2, while red for trivial gap with $\sum C_s^m = 0$. The first two gaps are always topologically nontrivial, while the third and fourth gaps change their topology twice due to mode exchanges. The color discontinues for the fourth gap near $\zeta_0 = 0.93$ as it is a topologically nontrivial partial gap.

The kernel on the topological phase transition of the third gap is the exchange between modes at high symmetry k-points. We plot the evolution of total spin Chern number and visualize the mode evolutions at Γ and M points near the critical transition points. In Fig. 3(a), when $\zeta_0$ reaches 0.816, two modes at Γ point with different $C_6$ eigenvalues $\eta^+$ are accidentally degenerated [red solid circle in Fig. 3(b)]. After the transition point, they start to exchange the sequence, leading that $\sum C_s^m$ should change from 0 to -2 [the left jump in Fig. 3(d)]. Similar phase transition occurs at $\zeta_0 = 0.858$ but at M point instead. Modes with different $C_2$ eigenvalues $\gamma^+$ exchange and thus the value of $\sum C_s^m$ has another jump in Fig. 3(d). The evolution of the fourth gap is similar to that of the third gap, except that the second transition is caused by mode exchange at K point.

However, there is no full band gap between two transition critical points (color discontinues in Fig. 2). Instead, it is a topologically nontrivial partial gap characterized by $\sum C_s^m = 3$.

Although the above discussions are focused on hexagonal bi-isotropic photonic crystals, the group-theory-based method of topological invariant determination is general. It can be also applied to gyromagnetic (gyroelectric) photonic crystals [8-11], magnetic photonic crystal slabs [31], uniaxial metacrystal waveguides [18], or piezoelectric and piezomagnetic superlattices [20]. We also demonstrate that the topological invariants retrieved by the group theory are consistent with the edge dispersion of bi-isotropic photonic crystals with square and honeycomb lattices [29].

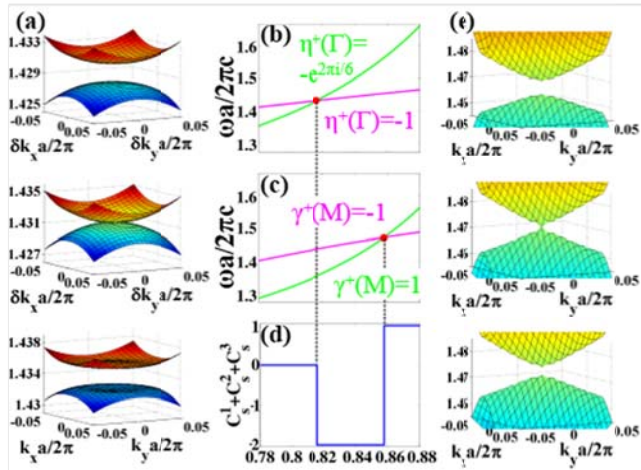

FIG. 3 (color online). Illustration on mode exchanges for the bands neighboring the third gap, showing its closing and reopening. (a) Eigen-frequency surfaces near Γ point and (b) evolutions at Γ point in the vicinity of $\zeta_0 = 0.816$. Red circle in (b) is the intersecting point between two modes with the $C_6$ eigenvalues ($\eta^+$) of $-e^{2\pi i/6}$ and -1. (c) & (e) Same as (b) & (a) but for the case of M point. (d) Schematic of the topological invariant $\sum C_s^m$ as a function of $\zeta_0$. Topological phase transition happens twice as $\sum C_s^m$ takes two jumps of -2 and 3 due to mode exchange at Γ and M points. Dashed lines are used to guide eyes.

The ε/µ matching condition is hard to fulfill in realistic systems, one question to be answered is that how robust when there is somewhat mismatching. To clarify the problem, we construct the spin edge states below light cone at the boundary between air and photonic topological insulator (PTI) of $(\varepsilon_r, \mu_r, \zeta_0)$ in the help of material generalizations [29]. We first start from the ideal ε/µ matching case of edge between air and the PTI with $(3, 3, 0.84*3)$. One-way spin-up edge state exists in the first gap and is evanescent in air. Figure 4(a) shows that the power flow can turn around the defects (either inserting insulating block or three missing rods) and keep moving rightwards. High transmittance is achieved in both two defects. Next, we consider a reasonable case of the boundary between the trivial air and the nontrivial PTI with $(9,1,0.84*3)$. The topological differences ensure the presence of edge states. The excited edge states can still pass through the two kinds of defects [Fig. 4(b)], although the rightward flow of light suffers from reflection. The loss of transmitted energy is caused by the backscattering due to ε/µ mismatching and the scattering into the air. As a fair comparison, we also did the case of the boundary between two POIs, i.e. the air and the ordinary photonic crystal of dielectric rod ($\varepsilon = 9$) in air background. The energy of transverse magnetic polarization totally reflects and there is no transmitted waves on the right side of the boundary [Fig. 4(c)]. The distinct results between Fig. 4(b) and Fig. 4(c) shows that the topological difference is a key factor for the design of fault-tolerant channel, waveguide or other devices.

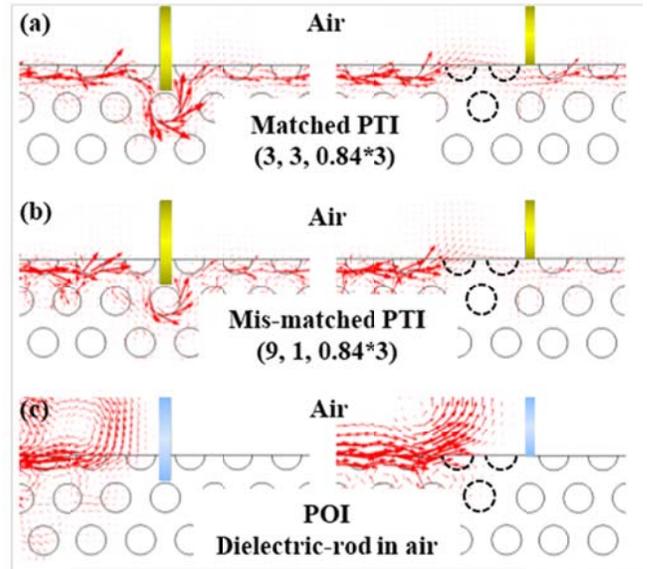

FIG. 4 (color online). Robust transport for an insulating block (rectangle, left column) and the missing rods (dash circle, right column). (a) ε/µ matching edge between air and the photonic topological insulator (3, 3, 0.84*3) along the ΓK direction. High transmittance is observed for both cases. (b) ε/µ mismatching edge between air and another photonic topological insulator (9, 1, 0.84*3). High transmittance is observed for the block. While for the case of missing rods, the waves have little air-loss but can still guide rightwards along the edge. (c) Edge between air and photonic ordinary insulator of dielectric rods in air background. None of transmitted wave can pass through either defect and the waves reflects completely. Red arrows are simulated power flows. Yellow block is material of $\varepsilon_r = \mu_r = \mathrm{diag}(1,1,-10000)$, while blue block is perfect electric conductor. Note that the insulating blocks on the right do not enter into the crystals and remove the air scattering to make figures clear.

In summary, we demonstrate that calculation of the topological invariant in bi-isotropic photonic crystals can be predigested by employing group theory. After retrieving pseudo-spin polarized fields, spin Chern number can be calculated, up to a multiple of n, by evaluating the eigenvalues of symmetry operators at high symmetry k-points. Total spin Chern number accurately predicts topological characters of photonic edge states and photonic band gaps. Nontrivial phase transition, which is caused by mode exchange, is found when the magnetoelectric coupling is large enough to induce the jump of total spin Chern number. Light transport of pseudo-spin polarized state is discussed in the ε/μ matching or mismatching edges.

This work is supported by grants of NSFC (11274396) and 973 program (2014CB931700). JWD is also supported by Guangdong Distinguished Young Scholar and Program for New Century Excellent Talents in University. This work is in part supported by supercomputer "Tianhe-II" and "Nanfang-I".